\documentclass[12pt]{article}
\usepackage{amsmath,amssymb}

\begin{document}

\begin{center}

{\Large \bf
A derivation of the Breit equation from Barut's covariant formulation of electrodynamics in terms of direct interactions}
\vskip .6in

Domingo J. Louis-Martinez
\vskip .2in

Department of Physics and Astronomy,\\ University of British
Columbia\\Vancouver, Canada, V6T 1Z1 

martinez@physics.ubc.ca

\end{center}
 
\vskip 3cm
 
\begin{abstract}
We study Barut's covariant equations describing the electromagnetic interactions between $N$ spin-$1/2$ particles.
In the covariant formulation each particle is described by a Dirac spinor. 
It is assumed that the  interactions between the particles are not mediated by a bosonic field 
(direct interactions). 
Within this formulation, using the Lagrangian formalism, we derive
the  approximate (semirelativistic) Breit equation for two interacting
spin-$1/2$ particles. 
\end{abstract}

PCAS No. 03.30.+p, 03.50.-z, 03.65.Pm, 11.10.Ef.

It is well known that an isolated system of $N$ particles interacting
electromagnetically may be described, to terms of second order, 
by the Darwin Lagrangian \cite{darwin},\cite{landau}. 

At the quantum level one can describe a system of two spin-$1/2$
particles interacting electromagnetically, including second order terms,
by the Breit equation \cite{breit, dirac1}. Perturbative solutions of the Breit equation
give results for the energy spectrum of relativistic bound systems in agrement with experiments
\cite{bethe, kroli, fushchich}.

A derivation of the Breit equation from standard quantum field theory was given in \cite{bethe}. A more recent derivation in terms of
direct interactions between fermions was given in \cite{duviryak}. The derivation presented in \cite{duviryak} uses elements of quantum field
theory and requires employing an unconventional vacuum state.

In this paper we work within the theory proposed by Barut \cite{barut1,barut2}. This theory is an alternative approach to quantum field theory
and does not require second quantization. It is a theory of direct interaction between spin-$1/2$ particles. Each particle is described by a Dirac spinor. 
It is assumed that the electromagnetic interactions between the particles are not mediated by a bosonic field 
(direct interactions). 

Our goal in this paper is to derive the Breit equation 
directly from Barut's covariant equations describing the electromagnetic interactions between two spin-$1/2$ particles.

The Darwin Lagrangian \cite{darwin},\cite{landau}, describing a system of $N$ particles interacting
electromagnetically, to terms of second order, can be written as:

\begin{eqnarray}
L_{Darwin} & = & - c^{2}\sum\limits^{}_{a} m_{a}
+ \frac{1}{2} \sum\limits^{}_{a} m_{a} v_{a}^{2}
+ \frac{1}{8 c^{2}}\sum\limits^{}_{a} m_{a} v_{a}^{4}
- \frac{1}{2} \sum\limits^{}_{a} \sum\limits^{}_{b \neq a}
\frac{e_{a} e_{b}}{r_{ab}}
\nonumber\\
& & + \frac{1}{4 c^{2}}
\sum\limits^{}_{a} \sum\limits^{}_{b \neq a}
\frac{e_{a} e_{b}}{r_{ab}}
\left((\vec{v}_{a} \vec{v}_{b}) 
+(\vec{n}_{ab} \vec{v}_{a})(\vec{n}_{ab} \vec{v}_{b})\right),
\label{P1}
\end{eqnarray}

\noindent where $c$ is the speed of light, $m_{a}$ and $e_{a}$ are the mass and electric charge of
particle $a$ ($a = 1,2,...,N$), $\vec{v}_{a}$ its velocity, and
$\vec{r}_{ab} = \vec{r}_{a} - \vec{r}_{b}$ the relative position of
particle $a$ with respect to particle $b$. 
In (\ref{P1}), $\vec{n}_{ab} \equiv \frac{\vec{r}_{ab}}{r_{ab}}$.

The Darwin Lagrangian can be derived either from
Faraday-Maxwell's field theory of electrodynamics \cite{landau} or from
the relativistic action-at-a-distance theory of Wheeler and Feynman
\cite{anderson, domingo3}.

To terms of fourth order ($\frac{v^4}{c^4}$),
ignoring the radiation
effects (which for Faraday-Maxwell electrodynamics appear at third order)
the Lagrangian for an isolated
system of N particles interacting electromagnetically
was obtained in \cite{smoro}, \cite{ohta}.

At the quantum level a system of two spin-$1/2$
particles interacting electromagnetically, including second order terms,
can be described by the Breit equation \cite{breit, dirac1}:

\begin{eqnarray}
& & \left(i\hbar \frac{\partial}{\partial t}  + i\hbar c \vec{\alpha}_{1}\frac{\partial}{\partial \vec{r}_{1}} +
i\hbar c \vec{\alpha}_{2}\frac{\partial}{\partial \vec{r}_{2}} - m_{1}c^{2}\beta_{1} - m_{2}c^{2}\beta_{2}\right.\nonumber\\
& & \left. - \frac{e_{1} e_{2}}{|\vec{r}_{1} -\vec{r}_{2}|} \left(1 - \frac{1}{2}\left((\vec{\alpha}_{1}\vec{\alpha}_{2})
+ \frac{(\vec{\alpha}_{1}(\vec{r}_{1} - \vec{r}_{2}))}{|\vec{r}_{1} 
-\vec{r}_{2}|}\frac{(\vec{\alpha}_{2}(\vec{r}_{1} - \vec{r}_{2}))}{|\vec{r}_{1} -\vec{r}_{2}|}\right)\right)\right)
\Psi(t, \vec{r}_{1}, \vec{r}_{2}) = 0.
\label{P2}
\end{eqnarray}

\noindent where $\Psi(t, \vec{r}_{1}, \vec{r}_{2})$ is a two-body bilocal field defined as follows:

\begin{equation}
\Psi(t,\vec{r}_{1},\vec{r}_{2}) = \psi_{1}\left(t, \vec{r}_{1}\right) \otimes \psi_{2}\left(t, \vec{r}_{2}\right).
\label{P3}
\end{equation}

\noindent and $\vec{\alpha}$ and $\beta$ are Dirac's matrices\cite{bjorkendrell}, $\otimes$ is used to denote a tensor product,
and $\hbar$ is Planck's constant.

Equation (\ref{P2}) is consistent with the Darwin
Lagrangian (\ref{P1}) for $N = 2$ \cite{breit}.
It was obtained by following the simple recipe \cite{breit, 2bDirac}
of substituting the free terms in (\ref{P1}) by the left hand sides of the free Dirac equation for each particle,
and by substituting the velocities $\vec{v}_{a}$ by $c \vec{\alpha}_{a}$ 
in the interaction terms of (\ref{P1}).

Let us consider a system of $N$ distinguishable spin-$1/2$ particles interacting 
electromagnetically. In Barut's formulation each particle is described by a Dirac spinor $\psi_{a}(x)$ ($a=1,2,...,N$). 
It is assumed that the interactions between particles are not mediated by a bosonic field. The action functional for each individual 
particle can be written as\cite{barut1,barut2}:

\begin{equation}
S^{(a)} = \int d^{4}x \left(i\hbar \bar{\psi}_{a}(x) \gamma^{\mu} \partial_{\mu} \psi_{a}(x) - m_{a}c \bar{\psi}_{a}(x)\psi_{a}(x)\right)
- \frac{e_{a}}{c} \int d^{4}x A^{(a)}_{\mu}(x) \bar{\psi}_{a}(x) \gamma^{\mu} \psi_{a}(x),
\label{1}
\end{equation}

\noindent where,

\begin{equation}
A^{(a)}_{\mu}(x) = \sum\limits_{b \ne a}^{} e_{b} \int d^{4}y \delta\left((x - y)^{2}\right) \bar{\psi}_{b}(y) \gamma_{\mu} \psi_{b}(y).
\label{2}
\end{equation}

The argument of the Dirac delta function in (\ref{2}) is $(x - y)^{2} = \eta_{\mu\nu} (x^{\mu} - y^{\mu})(x^{\nu} - y^{\nu})$, where the metric
tensor $\eta_{\mu\nu} = diag(+1, -1, -1, -1)$. The Dirac matrices $\gamma^{\mu}$ are defined as usual\cite{bjorkendrell}: $\gamma^{0}=\beta$, 
$\vec{\gamma}=\beta\vec{\alpha}$.
In (\ref{1},\ref{2}) $\bar{\psi}_{a}(x) = \psi^{\dagger}_{a}(x) \gamma^{0}$.

The action functional for the whole system of $N$ interacting distinguishable spin-$1/2$ particles can be given in the form\cite{barut2}:

\begin{eqnarray}
S & = & \sum\limits_{a}^{}\int d^{4}x \left(i\hbar \bar{\psi}_{a}(x) \gamma^{\mu} \partial_{\mu} \psi_{a}(x) - m_{a}c \bar{\psi}_{a}(x)\psi_{a}(x)\right)\nonumber\\
& & - \frac{1}{2c} \sum\limits_{a}^{} \sum\limits_{b \ne a}^{} e_{a} e_{b}\int\int d^{4}x d^{4}y  \delta\left((x - y)^{2}\right)
 \bar{\psi}_{a}(x) \gamma^{\mu} \psi_{a}(x) \bar{\psi}_{b}(y) \gamma_{\mu} \psi_{b}(y).
\label{3}
\end{eqnarray}

The action (\ref{3}) is invariant under Lorentz transformations. At the classical level, (\ref{3}) corresponds to the action functional of
Wheeler-Feynman action-at-a-distance theory of electrodynamics\cite{wheeler, anderson, hoyle, barut}:

\begin{equation}
S = - \sum\limits^{}_{a} m_{a} c \int d s_{a} \left(\dot{z}_{a}^{2}\right)^{\frac{1}{2}} 
- \frac{1}{2 c}\sum\limits^{}_{a}  \sum\limits^{}_{b\neq a}
e_{a} e_{b}
\int \int d s_{a} d s_{b}
\delta\left(\left(z_{a} - z_{b}\right)^{2}\right) \left(\dot{z}_{a} \dot{z}_{b}\right).
\label{4}
\end{equation}

In (\ref{4}), $s_{a}=c\tau_{a}$, where $\tau_{a}$ is the proper time of particle $a$, $z_{a}^{\mu}(s_{a})$ is its world line in Minkowski spacetime,
and the 4-vector velocity $\dot{z}_{a}^{\mu} = \frac{d z_{a}^{\mu}}{d s_{a}}$. The Dirac delta function in (\ref{4}) accounts for the interactions 
propagating at the speed of light forward and backward in time.

The fully relativistic equations of motion corresponding to the action functional (\ref{4}) admit
exact circular solutions for any number of particles \cite{schild, domingo}.

From the action functional (\ref{3}) the covariant equations for the spin-$1/2$ particles can be obtained as follows\cite{barut2}:

\begin{equation}
i\hbar \gamma^{\mu} \partial_{\mu} \psi_{a}(x) - m_{a}c \psi_{a}(x)
- \frac{e_{a}}{c}  \sum\limits_{b \ne a}^{} e_{b}\int d^{4}y  \delta\left((x - y)^{2}\right)
\bar{\psi}_{b}(y) \gamma_{\mu} \psi_{b}(y) \gamma^{\mu} \psi_{a}(x) = 0,
\label{5}
\end{equation} 

\begin{equation}
i\hbar \partial_{\mu} \bar{\psi}_{a}(x) \gamma^{\mu}  + m_{a}c \bar{\psi}_{a}(x)
+ \frac{e_{a}}{c}  \sum\limits_{b \ne a}^{} e_{b}\int d^{4}y  \delta\left((x - y)^{2}\right)
\bar{\psi}_{b}(y) \gamma_{\mu} \psi_{b}(y) \bar{\psi}_{a}(x)\gamma^{\mu}  = 0.
\label{6}
\end{equation} 

From (\ref{5}, \ref{6}) it follows that:

\begin{equation}
\partial_{\mu}\left(\bar{\psi}_{a}(x) \gamma^{\mu} \psi_{a}(x)\right) = 0,
(a=1,...,N).
\label{7}
\end{equation}

These are the well known continuity equations\cite{bjorkendrell}.

We use the normalization conditions:

\begin{equation}
\int d^{3}\vec{x} \psi^{\dagger}_{a}(t,\vec{x})\psi_{a}(t,\vec{x}) = 1.
\label{8}
\end{equation}

In \cite{barut1, barut2} from (\ref{5}, \ref{6}) an exact covariant equation for the two-body bilocal field: 

\begin{equation}
\Phi(t,\vec{x},\vec{y}) = \psi_{1}\left(t, \vec{x}\right) \otimes \psi_{2}\left(t - \frac{|\vec{x}-\vec{y}|}{c}, \vec{y}\right)
\end{equation}

\noindent was obtained.

In this paper we are interested in obtaining an approximate (semirelativistic) equation for the two-body bilocal field (\ref{P3}):

\begin{equation}
\Psi(t,\vec{x},\vec{y}) = \psi_{1}\left(t, \vec{x}\right) \otimes \psi_{2}\left(t, \vec{y}\right).
\label{9}
\end{equation}

We will now derive the approximate (semirelativistic) Breit equation for the two-body bilocal field (\ref{9}) directly from Barut's
covariant equations describing the interactions between two spin-$1/2$ particles.

In order to obtain the Breit equation  for $\Psi(t,\vec{x},\vec{y})$ (\ref{9})
in the case of electrodynamics, 
we return to the action functional (\ref{3}) for $N=2$:

\begin{eqnarray}
S & = & \int d^{4}x \left(i\hbar \bar{\psi}_{1}(x) \gamma^{\mu} \partial_{\mu} \psi_{1}(x) - m_{1}c \bar{\psi}_{1}(x)\psi_{1}(x)\right)\nonumber\\
& + & \int d^{4}y \left(i\hbar \bar{\psi}_{2}(y) \gamma^{\mu} \partial_{\mu} \psi_{2}(y) - m_{2}c \bar{\psi}_{2}(y)\psi_{2}(y)\right)\nonumber\\
& & - \frac{e_{1} e_{2}}{c}  \int\int d^{4}x d^{4}y  \delta\left((x - y)^{2}\right)
\bar{\psi}_{1}(x) \gamma^{\mu} \psi_{1}(x) \bar{\psi}_{2}(y) \gamma_{\mu} \psi_{2}(y).
\label{10}
\end{eqnarray}

In (\ref{10}) we write the Dirac delta function:

\begin{equation}
\delta\left((x - y)^{2}\right) = \frac{1}{2|\vec{x}-\vec{y}|} \left(\delta(y^{0} - x^{0} + |\vec{x}-\vec{y}|) +
\delta(y^{0} - x^{0} - |\vec{x}-\vec{y}|)\right).
\label{11}
\end{equation}

Therefore, we can write the third term in (\ref{10}) as:

\begin{eqnarray}
& & - \frac{e_{1} e_{2}}{c}  \int\int d^{4}x d^{4}y  \delta\left((x - y)^{2}\right) 
\bar{\psi}_{1}(x) \gamma^{\mu} \psi_{1}(x) \bar{\psi}_{2}(y) \gamma_{\mu} \psi_{2}(y) = \nonumber\\
& & - \frac{e_{1} e_{2}}{2} \int dt \int d^{3}\vec{x} \int \frac{d^{3}\vec{y}}{|\vec{x}-\vec{y}|}
\left(\bar{\psi}_{2}\left(t - \frac{|\vec{x}-\vec{y}|}{c},\vec{y}\right) \gamma_{\mu} \psi_{2}\left(t- \frac{|\vec{x}-\vec{y}|}{c},\vec{x}\right) +\right.\nonumber\\
& &\left. \bar{\psi}_{2}\left(t + \frac{|\vec{x}-\vec{y}|}{c},\vec{y}\right) \gamma_{\mu} \psi_{2}\left(t + \frac{|\vec{x}-\vec{y}|}{c},\vec{x}\right)\right) 
\bar{\psi}_{1}(t,\vec{x}) \gamma^{\mu} \psi_{1}(t,\vec{x})
\label{12}
\end{eqnarray}

Using the approximate expressions:

\begin{eqnarray}
\bar{\psi}_{2}\left(t - \frac{|\vec{x}-\vec{y}|}{c},\vec{y}\right) \gamma_{\mu} \psi_{2}\left(t- \frac{|\vec{x}-\vec{y}|}{c},\vec{x}\right)
& \approx & \bar{\psi}_{2}(t,\vec{y}) \gamma_{\mu} \psi_{2}(t,\vec{y}) 
- \frac{|\vec{x}-\vec{y}|}{c}\frac{\partial}{\partial t}\left(\bar{\psi}_{2}(t,\vec{y}) \gamma_{\mu} \psi_{2}(t,\vec{y})\right)\nonumber\\
& & +\frac{|\vec{x}-\vec{y}|^{2}}{2c^{2}}\frac{\partial^{2}}{\partial t^{2}}\left(\bar{\psi}_{2}(t,\vec{y}) \gamma_{\mu} \psi_{2}(t,\vec{y})\right)
\label{13}
\end{eqnarray}

\begin{eqnarray}
\bar{\psi}_{2}\left(t + \frac{|\vec{x}-\vec{y}|}{c},\vec{y}\right) \gamma_{\mu} \psi_{2}\left(t + \frac{|\vec{x}-\vec{y}|}{c},\vec{x}\right)
& \approx & \bar{\psi}_{2}(t,\vec{y}) \gamma_{\mu} \psi_{2}(t,\vec{y}) 
+ \frac{|\vec{x}-\vec{y}|}{c}\frac{\partial}{\partial t}\left(\bar{\psi}_{2}(t,\vec{y}) \gamma_{\mu} \psi_{2}(t,\vec{y})\right)\nonumber\\
& & +\frac{|\vec{x}-\vec{y}|^{2}}{2c^{2}}\frac{\partial^{2}}{\partial t^{2}}\left(\bar{\psi}_{2}(t,\vec{y}) \gamma_{\mu} \psi_{2}(t,\vec{y})\right)
\label{14}
\end{eqnarray}

and substituting (\ref{13},\ref{14}) in (\ref{12}) we obtain:

\begin{eqnarray}
& & - \frac{e_{1} e_{2}}{c}  \int\int d^{4}x d^{4}y  \delta\left((x - y)^{2}\right) 
\bar{\psi}_{1}(x) \gamma^{\mu} \psi_{1}(x) \bar{\psi}_{2}(y) \gamma_{\mu} \psi_{2}(y) \approx \nonumber\\
& & - e_{1} e_{2} \int dt \int d^{3}\vec{x} \int \frac{d^{3}\vec{y}}{|\vec{x}-\vec{y}|}
\bar{\psi}_{1}(t,\vec{x}) \gamma^{\mu} \psi_{1}(t,\vec{x}) \bar{\psi}_{2}(t,\vec{y}) \gamma_{\mu} \psi_{2}(t,\vec{y})\nonumber\\ 
& & -\frac{e_{1}e_{2}}{2c^{2}} \int dt \int d^{3}\vec{x} \int d^{3}\vec{y} |\vec{x}-\vec{y}|
\bar{\psi}_{1}(t,\vec{x}) \gamma^{\mu} \psi_{1}(t,\vec{x}) \frac{\partial^{2}}{\partial t^{2}}\left(\bar{\psi}_{2}(t,\vec{y}) \gamma_{\mu} \psi_{2}(t,\vec{y})\right)
\label{15}
\end{eqnarray}

Integrating the second term in (\ref{15}) by parts and recalling that $(\gamma^{0})^{2} = I, \gamma^{0}\vec{\gamma} = \vec{\alpha}$ we find:

\begin{eqnarray}
& & - \frac{e_{1} e_{2}}{c}  \int\int d^{4}x d^{4}y  \delta\left((x - y)^{2}\right) 
\bar{\psi}_{1}(x) \gamma^{\mu} \psi_{1}(x) \bar{\psi}_{2}(y) \gamma_{\mu} \psi_{2}(y) \approx \nonumber\\
& & - e_{1} e_{2} \int dt \int d^{3}\vec{x} \int \frac{d^{3}\vec{y}}{|\vec{x}-\vec{y}|}
\psi^{\dagger}_{1}(t,\vec{x})  \psi_{1}(t,\vec{x}) \psi^{\dagger}_{2}(t,\vec{y}) \psi_{2}(t,\vec{y})\nonumber\\ 
& & + e_{1} e_{2} \int dt \int d^{3}\vec{x} \int \frac{d^{3}\vec{y}}{|\vec{x}-\vec{y}|}
\psi^{\dagger}_{1}(t,\vec{x})  \vec{\alpha} \psi_{1}(t,\vec{x}) \psi^{\dagger}_{2}(t,\vec{y}) \vec{\alpha}\psi_{2}(t,\vec{y})\nonumber\\ 
& & +\frac{e_{1}e_{2}}{2c^{2}} \int dt \int d^{3}\vec{x} \int d^{3}\vec{y} |\vec{x}-\vec{y}|
\frac{\partial}{\partial t}\left(\psi^{\dagger}_{1}(t,\vec{x}) \psi_{1}(t,\vec{x})\right) 
\frac{\partial}{\partial t}\left(\psi^{\dagger}_{2}(t,\vec{y}) \psi_{2}(t,\vec{y})\right)\nonumber\\
& & -\frac{e_{1}e_{2}}{2c^{2}} \int dt \int d^{3}\vec{x} \int d^{3}\vec{y} |\vec{x}-\vec{y}|
\frac{\partial}{\partial t}\left(\psi^{\dagger}_{1}(t,\vec{x}) \vec{\alpha} \psi_{1}(t,\vec{x})\right) 
\frac{\partial}{\partial t}\left(\psi^{\dagger}_{2}(t,\vec{y}) \vec{\alpha} \psi_{2}(t,\vec{y})\right).
\label{16}
\end{eqnarray}

Using the continuity equations (\ref{7}), the third term in (\ref{16}) can be rewritten as:

\begin{eqnarray}
& & \frac{e_{1}e_{2}}{2c^{2}} \int dt \int d^{3}\vec{x} \int d^{3}\vec{y} |\vec{x}-\vec{y}|
\frac{\partial}{\partial t}\left(\psi^{\dagger}_{1}(t,\vec{x}) \psi_{1}(t,\vec{x})\right) 
\frac{\partial}{\partial t}\left(\psi^{\dagger}_{2}(t,\vec{y}) \psi_{2}(t,\vec{y})\right)\nonumber\\
& \approx & \frac{e_{1}e_{2}}{2} \int dt \int d^{3}\vec{x} \int d^{3}\vec{y} |\vec{x}-\vec{y}|
\frac{\partial}{\partial \vec{x}}\left(\psi^{\dagger}_{1}(t,\vec{x}) \vec{\alpha}\psi_{1}(t,\vec{x})\right) 
\frac{\partial}{\partial \vec{y}}\left(\psi^{\dagger}_{2}(t,\vec{y}) \vec{\alpha}\psi_{2}(t,\vec{y})\right).
\label{17}
\end{eqnarray}

Integrating by parts and taking into account that

\begin{equation}
\frac{\partial}{\partial \vec{y}}\left(|\vec{x} - \vec{y}|\right) = - \frac{\vec{x} - \vec{y}}{|\vec{x} - \vec{y}|}
\label{18}
\end{equation}

\noindent we obtain:

\begin{eqnarray}
& & \frac{e_{1}e_{2}}{2c^{2}} \int dt \int d^{3}\vec{x} \int d^{3}\vec{y} |\vec{x}-\vec{y}|
\frac{\partial}{\partial t}\left(\psi^{\dagger}_{1}(t,\vec{x}) \psi_{1}(t,\vec{x})\right) 
\frac{\partial}{\partial t}\left(\psi^{\dagger}_{2}(t,\vec{y}) \psi_{2}(t,\vec{y})\right)\nonumber\\
& \approx & \frac{e_{1}e_{2}}{2} \int dt \int d^{3}\vec{x} \int d^{3}\vec{y} 
\frac{\partial}{\partial \vec{x}}\left(\psi^{\dagger}_{1}(t,\vec{x}) \vec{\alpha}\psi_{1}(t,\vec{x})\right) 
\psi^{\dagger}_{2}(t,\vec{y}) \frac{\left(\vec{\alpha}(\vec{x} -\vec{y})\right)}{|\vec{x}-\vec{y}|}\psi_{2}(t,\vec{y}).
\label{19}
\end{eqnarray}

Performing a second integration by parts and taking into account that:

\begin{equation}
\frac{\partial}{\partial \vec{x}}\left(\frac{1}{|\vec{x} - \vec{y}|}\right) = - \frac{\vec{x} - \vec{y}}{|\vec{x} - \vec{y}|^{3}},
\label{20}
\end{equation}

\noindent we obtain:

\begin{eqnarray}
& & \frac{e_{1}e_{2}}{2c^{2}} \int dt \int d^{3}\vec{x} \int d^{3}\vec{y} |\vec{x}-\vec{y}|
\frac{\partial}{\partial t}\left(\psi^{\dagger}_{1}(t,\vec{x}) \psi_{1}(t,\vec{x})\right) 
\frac{\partial}{\partial t}\left(\psi^{\dagger}_{2}(t,\vec{y}) \psi_{2}(t,\vec{y})\right) \approx \nonumber\\
&  & - \frac{e_{1}e_{2}}{2} \int dt \int d^{3}\vec{x} \int \frac{d^{3}\vec{y}}{|\vec{x} - \vec{y}|} 
\psi^{\dagger}_{1}(t,\vec{x}) \vec{\alpha}\psi_{1}(t,\vec{x}) 
\psi^{\dagger}_{2}(t,\vec{y}) \vec{\alpha}\psi_{2}(t,\vec{y})\nonumber\\
& & + \frac{e_{1}e_{2}}{2} \int dt \int d^{3}\vec{x} \int \frac{d^{3}\vec{y}}{|\vec{x} - \vec{y}|}
\psi^{\dagger}_{1}(t,\vec{x}) \frac{\left(\vec{\alpha}(\vec{x} -\vec{y})\right)}{|\vec{x}-\vec{y}|}\psi_{1}(t,\vec{x}) 
\psi^{\dagger}_{2}(t,\vec{y}) \frac{\left(\vec{\alpha}(\vec{x} -\vec{y})\right)}{|\vec{x}-\vec{y}|}\psi_{2}(t,\vec{y}).
\label{21}
\end{eqnarray}

Substituting (\ref{21}) into (\ref{16}) we finally obtain:

\begin{eqnarray}
& & - \frac{e_{1} e_{2}}{c}  \int\int d^{4}x d^{4}y  \delta\left((x - y)^{2}\right) 
\bar{\psi}_{1}(x) \gamma^{\mu} \psi_{1}(x) \bar{\psi}_{2}(y) \gamma_{\mu} \psi_{2}(y) \approx \nonumber\\
& & - e_{1} e_{2} \int dt \int d^{3}\vec{x} \int \frac{d^{3}\vec{y}}{|\vec{x}-\vec{y}|}
\psi^{\dagger}_{1}(t,\vec{x})  \psi_{1}(t,\vec{x}) \psi^{\dagger}_{2}(t,\vec{y}) \psi_{2}(t,\vec{y})\nonumber\\ 
& & + \frac{e_{1} e_{2}}{2} \int dt \int d^{3}\vec{x} \int \frac{d^{3}\vec{y}}{|\vec{x}-\vec{y}|}
\psi^{\dagger}_{1}(t,\vec{x})  \vec{\alpha} \psi_{1}(t,\vec{x}) \psi^{\dagger}_{2}(t,\vec{y}) \vec{\alpha}\psi_{2}(t,\vec{y})\nonumber\\ 
& & +\frac{e_{1}e_{2}}{2} \int dt \int d^{3}\vec{x} \int \frac{d^{3}\vec{y}}{|\vec{x} - \vec{y}|}
\psi^{\dagger}_{1}(t,\vec{x}) \frac{\left(\vec{\alpha}(\vec{x} -\vec{y})\right)}{|\vec{x}-\vec{y}|}\psi_{1}(t,\vec{x}) 
\psi^{\dagger}_{2}(t,\vec{y}) \frac{\left(\vec{\alpha}(\vec{x} -\vec{y})\right)}{|\vec{x}-\vec{y}|}\psi_{2}(t,\vec{y})\nonumber\\
& & -\frac{e_{1}e_{2}}{2c^{2}} \int dt \int d^{3}\vec{x} \int d^{3}\vec{y} |\vec{x}-\vec{y}|
\frac{\partial}{\partial t}\left(\psi^{\dagger}_{1}(t,\vec{x}) \vec{\alpha} \psi_{1}(t,\vec{x})\right) 
\frac{\partial}{\partial t}\left(\psi^{\dagger}_{2}(t,\vec{y}) \vec{\alpha} \psi_{2}(t,\vec{y})\right).
\label{22}
\end{eqnarray}

If we assume that the last term in (\ref{22}) can be neglected, then, substituting (\ref{22}) into (\ref{10}) we obtain the approximate
action functional for two spin-$1/2$ particles in the form:

\begin{eqnarray}
& & S \approx\nonumber\\
& & \int dt\int d^{3}\vec{x}\int d^{3}\vec{y} \psi^{\dagger}_{2}(t, \vec{y})\psi_{2}(t, \vec{y}) \psi^{\dagger}_{1}(t, \vec{x})
\left(i\hbar  \frac{\partial}{\partial t} \psi_{1}(t, \vec{x}) 
+ i\hbar c \vec{\alpha}\frac{\partial}{\partial \vec{x}}\psi_{1}(t, \vec{x})
- m_{1}c^{2} \beta\psi_{1}(t, \vec{x})\right) \nonumber\\
& & + \int dt \int d^{3}\vec{x} \int d^{3}\vec{y}\psi^{\dagger}_{1}(t, \vec{x})\psi_{1}(t, \vec{x})
\psi^{\dagger}_{2}(t, \vec{y})\left(i\hbar \frac{\partial}{\partial t} \psi_{2}(t, \vec{y}) 
+ i\hbar c \vec{\alpha}\frac{\partial}{\partial \vec{y}}\psi_{2}(t, \vec{y})
- m_{2}c^{2} \beta\psi_{2}(t, \vec{y})\right) \nonumber\\
& & - e_{1} e_{2} \int dt \int d^{3}\vec{x} \int \frac{d^{3}\vec{y}}{|\vec{x}-\vec{y}|}
\psi^{\dagger}_{1}(t,\vec{x})  \psi_{1}(t,\vec{x}) \psi^{\dagger}_{2}(t,\vec{y}) \psi_{2}(t,\vec{y})\nonumber\\ 
& & + \frac{e_{1} e_{2}}{2} \int dt \int d^{3}\vec{x} \int \frac{d^{3}\vec{y}}{|\vec{x}-\vec{y}|}
\psi^{\dagger}_{1}(t,\vec{x})  \vec{\alpha} \psi_{1}(t,\vec{x}) \psi^{\dagger}_{2}(t,\vec{y}) \vec{\alpha}\psi_{2}(t,\vec{y})\nonumber\\ 
& & +\frac{e_{1}e_{2}}{2} \int dt \int d^{3}\vec{x} \int \frac{d^{3}\vec{y}}{|\vec{x} - \vec{y}|}
\psi^{\dagger}_{1}(t,\vec{x}) \frac{\left(\vec{\alpha}(\vec{x} -\vec{y})\right)}{|\vec{x}-\vec{y}|}\psi_{1}(t,\vec{x}) 
\psi^{\dagger}_{2}(t,\vec{y}) \frac{\left(\vec{\alpha}(\vec{x} -\vec{y})\right)}{|\vec{x}-\vec{y}|}\psi_{2}(t,\vec{y}).
\label{23}
\end{eqnarray}

Following \cite{barut1, barut2}, we have multiplied the free terms in the action functional by the normalization conditions (\ref{8}). 
This can of course be done
without altering these terms (it is equivalent to multiplying each term by one).

In terms of the two-body bilocal spinor (\ref{9}) the action (\ref{23}) can simply be written as:

\begin{eqnarray}
S & \approx &
\int dt\int d^{3}\vec{x}\int d^{3}\vec{y} \Psi^{\dagger}(t, \vec{x}, \vec{y})
\left(i\hbar \frac{\partial}{\partial t}  + i\hbar c \vec{\alpha}_{1}\frac{\partial}{\partial \vec{x}}
+ i\hbar c \vec{\alpha}_{2}\frac{\partial}{\partial \vec{y}} - m_{1}c^{2}\beta_{1} - m_{2}c^{2}\beta_{2}\right.\nonumber\\
& & \left. - \frac{e_{1} e_{2}}{|\vec{x} -\vec{y}|} \left(1 - \frac{1}{2}\left((\vec{\alpha}_{1}\vec{\alpha}_{2})
+ \frac{(\vec{\alpha}_{1}(\vec{x} - \vec{y}))}{|\vec{x} -\vec{y}|}\frac{(\vec{\alpha}_{2}(\vec{x} - \vec{y}))}{|\vec{x} -\vec{y}|}\right)\right)\right)
\Psi(t, \vec{x}, \vec{y})
\label{24}
\end{eqnarray}

From this action functional, one can immediately obtain Breit's equation for electrodynamics in the form: 

\begin{eqnarray}
& & \left(i\hbar \frac{\partial}{\partial t}  + i\hbar c \vec{\alpha}_{1}\frac{\partial}{\partial \vec{x}} +
i\hbar c \vec{\alpha}_{2}\frac{\partial}{\partial \vec{y}} - m_{1}c^{2}\beta_{1} - m_{2}c^{2}\beta_{2}\right.\nonumber\\
& & \left. - \frac{e_{1} e_{2}}{|\vec{x} -\vec{y}|} \left(1 - \frac{1}{2}\left((\vec{\alpha}_{1}\vec{\alpha}_{2})
+ \frac{(\vec{\alpha}_{1}(\vec{x} - \vec{y}))}{|\vec{x} -\vec{y}|}\frac{(\vec{\alpha}_{2}(\vec{x} - \vec{y}))}{|\vec{x} -\vec{y}|}\right)\right)\right)
\Psi(t, \vec{x}, \vec{y}) = 0.
\label{25}
\end{eqnarray}

\noindent which coincides with (\ref{P2}).

We have presented a very simple derivation of the Breit equation from Barut's formulation of electrodynamics in terms of direct interactions between
fermions. In this formulation the spinors describing the spin-$1/2$ particles are not operators. This is reflected at every step of our derivation,
which proceeds completely within the realm of the Lagrangian formalism of relativistic wave equations.

Since it is known that the Breit equation gives results, up to terms of order $\alpha^{4}$, for the energy spectrum of relativistic bound 
systems (such as positronium) in agreement with
experiments\cite{kroli, fushchich}, our proof that the Breit equation follows from 
Barut's formulation of electrodynamics (in terms of direct interactions) lends further support to Barut's approach.


\begin{thebibliography}{99}


\bibitem{darwin} C.G. Darwin, Phil. Mag. {\bf 39}, 537 (1920).
\bibitem{landau} L. D. Landau and E. M. Lifshitz, {\it The Classical Theory
of Fields} (4th revised English edition), Butterworth- Heinenann, Oxford (1996).
\bibitem{breit}G. Breit, Phys. Rev. {\bf 34}, 553 (1929).
\bibitem{dirac1} P.A.M. Dirac, R. Peierls and M.H.L. Pryce, Proc. Cambridge Phil. Soc. {\bf 35}, 186 (1939).
\bibitem{bethe} H.A. Bethe and E.E. Salpeter,
{\it  Quantum Mechanics of One and Two Electron Atoms}
(Springer, Berlin, 1957).
\bibitem{kroli} W. Krolikowski, Acta Phys. Pol. {\bf B12}, 891 (1981).
\bibitem{fushchich} V.I. Fushchich and A.G. Nikitin, {\it Symmetries of Equations of Quantum Mechanics} (Allerton Press, 1994).
\bibitem{duviryak} A. Duviryak and J.W. Darewych, CEJP, 3(4), 467 (2005).
\bibitem{barut1}A.O. Barut, Lecture Notes in Physics, Vol. {\bf 180}, p. 332 (Springer-Verlag, Berlin, 1983).
\bibitem{barut2}A.O. Barut, J. Math. Phys. {\bf 32}, 1091 (1991).
\bibitem{anderson} J.L. Anderson, {\it Principles of Relativity Physics}
(Academic Press, New York, 1967).
\bibitem{domingo3} D.J. Louis-Martinez, Phys. Lett. A{\bf 364}, 93 (2007).
\bibitem{smoro} V.N. Golubenkov and Ya. A. Smorodinskii, Zh. Eksp. Teor. Fiz. {\bf 31}, 330 (1956)
[Sov. Phys. JETP {\bf 4}, 442 (1957)].
\bibitem{ohta} T. Ohta and T. Kimura, J. Math. Phys. {\bf 34}, 4655 (1993).
\bibitem{bjorkendrell} J.D. Bjorken and S.D. Drell, {\it Relativistic Quantum Mechanics} (McGraw-Hill, New York, 1964).
\bibitem{2bDirac}P. Van Alstine and H. Crater, Found. Phys. {\bf 27}, 67
(1997).
\bibitem{wheeler}J. A. Wheeler and R. P. Feynman, Rev. Mod. Phys. {\bf 17}, 157 (1945); {\bf 21}, 425 (1949).
\bibitem{hoyle} F. Hoyle and J.V. Narlikar,
{\it Lectures on Cosmology and Action at a Distance Electrodynamics},
(World Scientific, 1996).
\bibitem{barut} A. O. Barut, {\it Electrodynamics and Classical Theory of
Fields and Particles}, (Dover Publications Inc, New York, 1980).
\bibitem{schild} A. Schild, Phys. Rev. {\bf 131}, 2762 (1963).
\bibitem{domingo} D.J. Louis-Martinez, Phys. Lett. A{\bf 320}, 103  (2003).


\end{thebibliography}
\end{document}